# Hilbert-Huang Dönüşümü ve Derin Öğrenme Kullanarak Akciğer Seslerinde Astım Teşhisi

# The Diagnosis of Asthma using Hilbert-Huang Transform and Deep Learning on Lung Sounds

Gökhan Altan[1,*], Yakup Kutlu[2], Adnan Özhan Pekmezci[3], Serkan Nural[3]
[1,*]Enformatik ABD, Mustafa Kemal Üniversitesi, Hatay, Türkiye
gokhan_altan@hotmail.com
[2]Bilgisayar Mühendisliği Bölümü, İskenderun Teknik Üniversitesi, Hatay, Türkiye
[3]Antakya Devlet Hastanesi, Hatay, Türkiye

*Özetçe—* Akciğer oskültasyonu stetoskop kullanılarak nefes alma ve nefes verme süreçlerinde hava yollarında meydana gelen sesleri kullanarak çeşitli solunum rahatsızlıklarının teşhisinde kullanılan en etkili ve olmazsa olmaz bir yöntemdir. Bu çalışmada göğüs ve sırtta 12 farklı bölgeden kaydedilen akciğer seslerine Hilbert-Huang dönüşümü uygulanarak elde edilen yeni formdaki içsel mod fonksiyonlarından elde edilen istatistiksel öznitelikler hesaplanmıştır. Derin İnanç Ağları (DİA) kullanılarak astım ve sağlıklı akciğer seslerinin sınıflandırılmasını gerçekleştirmiştir. Çift gizli katmanlı DİA sınıflandırıcı modeli 5 parçalı çapraz doğrulama yöntemiyle test edilmiştir. Önerilen DİA modeli astımlı ve sağlıklı bireylerin akciğer seslerinin frekans-zaman analiziyle %84.61 genel başarım, %85.83 hassasiyet ve %77.11 belirlilikle ayrıştırmıştır.

*Anahtar Kelimeler—Derin öğrenme; akciğer oskültasyonu; Hilbert-Huang Dönüşümü; Derin inanç ağları; Astım; Hırıltı.*

*Abstract—* Lung auscultation is the most effective and indispensable method for diagnosing various respiratory disorders by using the sounds from the airways during inspirium and exhalation using a stethoscope. In this study, the statistical features are calculated from intrinsic mode functions that are extracted by applying the Hilbert-Huang Transform to the lung sounds from 12 different auscultation regions on the chest and back. The classification of the lung sounds from asthma and healthy subjects is performed using Deep Belief Networks (DBN). The DBN classifier model with two hidden layers has been tested using 5-fold cross validation method. The proposed DBN separated lung sounds from asthmatic and healthy subjects with high classification performance rates of 84.61%, 85.83%, and 77.11% for overall accuracy, sensitivity, and selectivity, respectively using frequency-time analysis.

*Keywords—Deep Learning; Lung Auscultation, Hilbert-Huang Transform; Deep Belief Networks; Asthma; Wheezing.*

## I. GİRİŞ

Bilgisayar destekli medikal tanı sistemleri (BDTS) günümüz teknolojisinde klinik işlemlerin gerçekleştirilmesi, çeşitli testlerin yapılması, yapılan testlerin yorumlanması ve hekimlere uzmanlıkları doğrultusunda detaylı ve belirleyici analiz bilgilerinin aktarılması hususunda sıklıkla kullanılan yöntem ve teknikler içerisindedir.

İnsanların sağlık kolunda yaşam kalitelerini artırmanın en önemli faktörlerin başında sağlıklı bir toplum oluşturma, hastalıkları ortadan kaldırma veya hastalıkları kontrol altında tutabilme kriterleri yer almaktadır. Bireylerin yaşam kalitelerinin yükseltilmesi ve sağlık konusuna en fazla etki eden faktörlerin belirlenmesi ve düzenlenmesi devlet politikalarında önemli yer tutmalıdır. Dünya Sağlık Örgütü'nün raporları göz önünde bulundurulduğunda dünyada en ölümcül 5 hastalık listesinde 3 adet solunum rahatsızlığı yer almaktadır [1]. Solunum rahatsızlıkları hava kirliliği, kalıtsal sebepler, tütün kullanımı veya sigara dumanına maruz kalma, yaş, cinsiyet, ırk, enfeksiyonlar, mevsimsel faktörler, coğrafik koşullar, mesleksel faktörler gibi çevresel etmenlere bağlı olarak meydana gelen bozukluklardır. En yaygın rastlanan solunum rahatsızlıkları kronik olarak gerçekleştirilen astım, bronşit, kronik obstrüktif akciğer hastalığı (KOAH), alt solunum yolları enfeksiyonu olarak kayıtlara geçmiştir [2]. En iyi şartları sağlayarak yaşam kalitesini ve en iyi sağlık düzeyini sürdürebilmenin yanında hastalıklar için tanı ve erken tanı sistemlerinin oluşturulması da hastalığa doğrudan ve hızlı müdahale için önem taşımaktadır [3].

Oskültasyon, iç organların stetoskop yardımıyla dinlenmesi işlemine verilen genel isimdir. Oskültasyon işlemi göğüs, kalp, mide ve bağırsak hastalıkları için sıklıkla kullanılan bir tanı yöntemidir. En önemli karakteristikleri ucuz, etkili, kolay kullanılabilir ve detaylı inceleme yapabilme imkânı sağlayan bir yöntem olmasıdır. Göğüs hastalıklarında teknolojinin sürekli ve hızla gelişmesine rağmen en yaygın kullanılan ve olmazsa olmaz tanı koyma yöntemi hala oskültasyondur [4]. Oskültasyon işlemiyle elde edilen kalp ve akciğer sesleri kardiyak bozuklukların, pulmoner rahatsızlıkların ve kardiyo-pulmoner rahatsızlıkların tanısını yapmada kullanılmaktadır. Özellikle akciğer oskültasyonu ile elde edilen sesler normal ve anormal olarak sınıflandırılmasıyla birlikte, anormal solunum sesleri farlı türde tıkayıcı ve esneklik bozucu hastalıklar sebebiyle hırıltı, çıtırtı sesler, stridor olarak sınıflandırılmaktadır [2], [5]. Akciğer seslerindeki bu anormal sesler hastalığın türüne göre farklılık göstermekte ve devamlılık süreleri değişkenlik göstermektedir [5].

Astım, kişinin ortamdaki çeşitli alerjenler, sigara dumanı, hava kirliliği, soğuk havaya maruz kalma gibi tetikleyici etmenlerle karşı karşıya kalması sonucu bronş adı verilen havayollarının daralması sonucu oluşan bir solunum rahatsızlığıdır [6], [7]. Astım sahip olduğu evreye göre belirli dönemlerde ataklarla kendini yoğun bir şekilde hissettiren kronik bir akciğer hastalığıdır. Atak dönemleri haricinde çoğunlukla hastalara aşırı bir rahatsızlık vermeyen rahatsızlık atak sırasında nefes darlığı, öksürük, hırıltılı solunum, göğüste sıkışıklık, nefes alamama ve bu durumlardan kaynaklı panik hissi gibi belirtilerle kendini gösterir [3], [7]. Hastalığın belirtilerinin şiddeti hastadan hastaya, hastanın aynı zamanda farklı kronik rahatsızlıklar geçiriyor olmasına ve hastalığın evresine göre farklılıklar gösterebilir. Kalıtsal özelliklere bağlı olarak da gelişebildiği için yalnızca yetişkinler için değil çocuklarda da sıklıkla tanısı konulan bir hastalıktır [8]. Bu açıdan tanısının erken konulması ve kontrol altına alınması her yaşı ilgilendiren ve bu alanda yapılması gereken araştırmalar olduğunu açıkça ortaya koyan bir alandır. Astım hastalarında duyulan akciğer sesleri hırıltı (wheezing) olarak geçen ve normal seslerin bazı bölümlerinde beklenmedik ve devamlı olarak gelen inflematuar ses niteliğine sahip anormal sestir. Akciğerdeki hava çeperlerinin daralması ve dolayısıyla hava yollarının daralması ile hava geçişi ve basınç değişimleri sırasında hırlama türü bir ses oluşmasına sebep olur. Amerikan Göğüs topluluğunun (American Thoratic Society) süreklilik hırıltının süresini tanımlamaktadır. 250ms ve üzeri hırıltılar astım ve KOAH hastalığını tanımlama da kullanıldığı gibi hırıltıların hissedildiği frekans 400 Hz ve yüksek frekanslardır. Hırıltılar frekans karakteristikleri ve müzikal karakteristikleri kullanılarak tespit edilebilmekte ve sınıflanabilmektedir [9]. Astım hastalarından çeşitli türlerde sensör ve mikrofonlarla kaydedilen oskültasyon sesleri üzerinde frekans ve zaman domeninde yapılan analizlerle hırıltı seslerinin otomatik tespit edilmesini sağlayan sistemler oluşturulmuştur. Bu çalışmalar içinde Jane vd. [10] özbağlanımlı modellemeyle (Autoregressive Model) güç spektrum yoğunlukları ve frekans tepe değerlerini kullanarak, Hossain vd. [11] Hilbert dönüşümü, hızlı Fourier dönüşümü ve lineer regresyon analiziyle ortalama genlik değerleri, ortalama akış ve ortalama güç özniteliklerini çıkararak, Wisniewski vd. [12] Tager enerji operatörü, kısa-zamanlı Fourier dönüşümü değerlerinden elde ettikleri göğüs seslerinden hesapladıkları kürtosis, entropi, ortalama çaprazlama değerleri yardımıyla, Altan vd. hırıltı seslerinin üç boyutlu ikinci derece fark haritalarını bölütlenmesini derin öğrenme algoritmaları kullanarak kronik obstruktif akciğer rahatsızlığının seviyelerinin teşhisini [13], Ulukaya vd. [14] önerdikleri Q değişkeni ayarlanabilir dalgacık dönüşümünü baz alan oranlı genişleme dalgacık dönüşümüyle çıkardıkları özniteliklerle hırıltı seslerinin tespitini ve astım hastalığının analizini gerçekleştirmişlerdir. Literatürden de anlaşıldığı üzere çalışmalar genel olarak frekans bazlı analizler kullanmıştır. Özellikle hırıltı sesler belirli frekans değeri aralıklarında rastlanan sesler olduğu için özel dönüşüm modelleri ile yüksek başarımlar elde edilmiştir.

Bu çalışmada, astım hastalarından alınmış kalp ve akciğer oskültasyon seslerine Hilbert-Huang dönüşümü (HHD) uygulanarak elde edilen içsel mod fonksiyonlarının (İMF) istatistiksel öznitelikleri çıkarılacaktır. HHD sayesinde oskültasyon kayıtlarının zaman-frekans domeninde incelenmesi gerçekleştirilecek ve astım rahatsızlığının belirlenmesi için derin öğrenme bazlı bir matematiksel bir sınıflandırıcı modellenecektir. Bu çalışmayla literatürde yer alan astım ve hırıltı seslerinin belirlenmesi çalışmalarına alternatif bir yöntem oluşturulacak ve başarımların iyileştirildiği bir uygulamaya dönüştürülecektir. Sonraki bölümlerde kullanılan oskültasyon seslerinin nasıl alındığı, hastalık dağılımları, hasta seçiminde dikkat edilen durumlar, oskültasyon seslerinin HHD ile analizi, öznitelik çıkarma yöntemi, derin inanç ağları (DİA) sınıflandırıcı modelleme ve İMF bazlı başarımların paylaşılması gerçekleştirilecektir. Elde edilen sonuçların yorumlanması ve literatüre göre avantaj ve dezavantajları tartışılacaktır.

## II. METOT ve YÖNTEM

Bu bölümde analizlerde kullanılan oskültasyon seslerinin elde edilmesi aşamasında gerçekleştirilen senaryolar, oskültasyon için seçilen bölgeler ve bu bölgelerin özellikleri ve hasta seçiminde göz önünde bulundurulan kriterler, öznitelik çıkarmak için kullanılan frekans-zaman analizi, öznitelik çıkarmayı sağlayan istatistiksel analiz işlemleri, sınıflandırma modeli açıklanacaktır.

*A. Veritabanı*

Göğüs hastalıklarında teknolojinin sürekli ve hızla gelişmesine rağmen en yaygın kullanılan ve olmazsa olmaz tanı koyma yöntemi hala oskültasyondur. Oskültasyon seslerinin dijitalleştirilmesinde son yıllarda yapılan çalışmalarda farklı özelliklere sahip elektronik stetoskoplar [14]–[17] gibi normal fiziksel muayeneler sırasında çok fazla tercih edilmeyen medikal malzemeler kullanılmaktadır. Özellikle göğüs seslerinde meydana gelen çok kısa çıtırtılar ve hırıltıların duyulması için oldukça hassas cihazların kullanılması şarttır. Bu çalışmada Littmann 3200 Elektronik stetoskop kullanılmıştır. Bu cihaz çevredeki gürültüyü kritik vücut seslerini ortadan kaldırmadan ortalama 75% (-12dB) azaltmanızı sağlar. Son teknoloji süzme devresi kalp, akciğer ve diğer vücut seslerinin dinlenmesi için üç frekans tepki modu (Bell, Diagphram, Extended) bulundurur.

Göğüs oskültasyonu için standartlaştırılmış (CORSA) bölgeler olmasına karşın akciğerin belirli bölgeleri gerekli görülürse oskültasyona dâhil edilebilmektedir. Oskültasyon işlemi sırasında dış seslerden izole edilmiş, vücut ısısında ayarlanmış, ışık alan, hastanın kendini rahat hissedebileceği bir ortam kullanılmıştır. Gönüllüler oturur pozisyonda fiziksel muayeneden geçirilmiş ve ses oskültasyon kayıtları alınmıştır. Kayıtlar sağ ve sol aynı bölgelerden iki adet dijital stetoskop kullanılarak bir göğüs uzmanı tarafından senkronize olarak kaydedilmiştir. Gönüllünün kayıt süresince konuşmaması, insandan kaynaklı yapay sesler çıkarmaması istenmiştir. Kayıt başlatıldığında ilk 5sn içerisinde hastanın bir defa öksürmesi ve sonrasında doktorun kayıt bitti yönergesine kadar ağzından derin nefes alıp vermesi istenmiştir. Fiziksel muayene sırasında kullanılan Göğüs oskültasyon bölgeleri arka-üst akciğer (L1-R1), arka-orta akciğer (L2-R2), arka-alt akciğer (L3-R3), arka-kostofrenik açısı (L4-R4), ön-üst akciğer (L5-R5), ön-alt akciğer (L6-R6) olarak tercih edilmektedir. Akciğer oskültasyon bölgeleri Şekil 1'de görüldüğü gibi 6 farklı bölgeden sağ ve sol senkronize kayıt edilmek suretiyle 12 adet akciğer sesi elde edilmiştir. Solunum seslerinin toplanmasıyla ilgili detaylı bilgi [18] 'de sunulmuştur.

Astım rahatsızlığına ait oskültasyon ses kayıtları 4 erkek, 2 bayan olmak üzere yaşları 40 ile 65 arasında değişen 6 gönüllüden alınmıştır. Sağlıklı oskültasyon kayıtları, şuana kadar hiç sigara veya tütün ürünleri kullanmamış gönüllüler arasında seçilmiştir. Gönüllü popülasyonu oluşturulurken farklı meslek ve sosyal-ekonomik durumlara sahip olması, şuana kadar herhangi bir kronik akciğer rahatsızlığı geçmişinin olmamasına dikkat edilmiştir. KOAH ve astım rahatsızlığının kalıtımsal aktarımı göz önünde bulundurularak birinci dereceden yakınlarında astım ve KOAH bulunan bireyler gönüllü popülasyonuna dâhil edilmemiştir. Oskültasyon seslerinin analizi yapılırken 10 saniyelik kayıtlar kullanılmıştır.

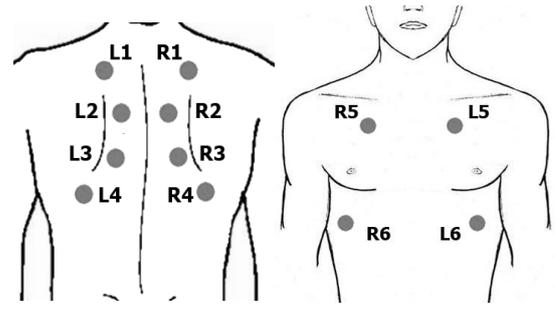

**Şekil 1.** Göğüs ve sırt oskültasyon bölgeleri

*B. Hilbert-Huang Dönüşümü*

HHD, lineer olmayan ve durağan olmayan sinyaller için anlık frekans değerlerinin hesaplanmasını sağlayan, NASA çalışanı Huang tarafından ortaya atılmış bir analiz yöntemidir. HHD durağan olmayan süreçlerden İMF elde etmeyi sağlayan son yıllarda sıkça kullanılan popüler bir ayrışım metodu olma özelliğine sahiptir. Fourier dönüşümü gibi teorik metotların aksine bir veri setine uygulanabilme özelliği sağlayan deneysel yaklaşımlar içeren bir dönüşümdür [19]. İki aşamalı bir süreçten meydana gelir. Birinci aşamasında sinyale ampirik kip ayrışımı (AKA) uygulanır ve sinyalin durumuna ve belirlenen ayırma algoritmasına göre İMF'ler ve bir artık sinyal elde edilir. İkinci aşamasında ise ayrıştırılan İMF'lerin Hilbert spektral analizi (HSA) elde edilir [20].

*C. Derin İnanç Ağları ve Performans Ölçme*

Derin öğrenme son 10 yılda yaygın olarak kullanılan ve bütünleştirilebilir matematiksel modellere sahip olmasından dolayı gelişimini hızlandırmış genellikle makine görmesi ve ses tanıma algoritmaları için sıklıkla kullanılan bir makine öğrenme algoritmasıdır. Derin öğrenme sayesinde çalışılan resim, video, EKG, vb. sinyal türlerinin daha detaylı analizinin kısa süreli eğitim algoritmalarıyla gerçekleştirilebilmektedir. DİA derin öğrenme algoritmaları içerisinde en yaygın kullanıma sahip olan, çok katmanlı gizli ve olasılıksal değişkenlerden oluşan olasılıklı geliştirilebilir bir sınıflandırıcı modelidir [21].

Genel başarım, hassasiyet ve belirlilik sınıflandırma veya kümeleme algoritmalarının performansının değerlendirilmesinde kullanılan test karakteristikleridir [22].

## III. DENEYSEL SONUÇLAR

BDTS üzerindeki yoğun çalışmalarla elde edilen gelişimler takip, teşhis, tanı ve analiz yetkinliklerinin yanı sıra hastalar hakkında elde edilen bulgu ve laboratuvar testlerinin saklanmasını sağlayan fonksiyonelliklerle kullanışlılığını artırmıştır. Elde edilen verilere her yerden kolay erişimin sağlanması amaçlarına göre modernize edilerek büyük bir hasta bulutu oluşturulması üzerindeki ihtiyaçları karşılayacak nitelikte sistemlere yönelik çalışmalarda büyük yollar kat edilmesini sağlamaktadır. BDTS'de hassasiyet ve güvenilirlik, hasta popülasyonunun çok geniş ve homojen dağılmadığı süreçlerin analizi ve testi için kullanılan hâkim bağımsız test karakteristikleridirler [22].

Hasta ve sağlıklı gönüllülerden alınan sinyallerin asıl etiketlenmesi sonucu elde edilen etiket vektörü ile modellenen sistem tarafından teşhis edilen hastalık veya sağlıklı olma durumunun karşıt çaprazlama sonrası elde edilen değerler değerlendirilmiş ve kullanılan frekans-zaman analiz yöntemi ve modellenen sınıflandırıcının performansı değerlendirilmiştir. Matematiksel modelleme ve öznitelik çıkarma süreci göz önünde bulundurularak astım hastalığı olan 5 gönüllü ile, şuana kadar hiç tütün ürünü kullanmamış ve herhangi bir kronik akciğer rahatsızlığı geçirmemiş 10 gönüllüden dijital stetoskop kullanılarak kaydedilen akciğer sesi sinyallerinden ele alınan sağ ve sol olmak üzere 12 kanaldan alınmış 10 saniyelik oskültasyon sesi kullanılmıştır. Her gönüllüden 12 kanal olmak üzere 180 adet akciğer oskültasyon sesi analizlerde kullanılmıştır. Akciğer oskültasyon sinyallerine AKA işlemi sonrasında IMF'ler çıkarılmıştır. Oskültasyon seslerinden elde edilen İMF sayıları sinyalin durumu ve frekans dağılımına göre değişiklik gösterebildiği için 5 ile 7 arasında değişen İMF'ler elde edilmiştir. Elde edilen her IMF'ye Hilbert dönüşümü uygulanarak HHD süreci tamamlanmıştır. AKA ile zaman domeninde, Hilbert dönüşüm ile frekans domeninde analiz edilen sinyalden yeni formdaki sinyaller çıkarılmıştır. HHD sonucu elde edilen İMF'lerden artık olarak kabul edilen son İMF'ler hariç tüm İMF'lerden ortalama, medyan, standart sapma, maksimum, minimum, varyans, sık tekrar eden değer (mod), korelasyon katsayısı gibi istatistiksel öznitelikler, basıklık (kurtosis), moment, kümülant gibi yüksek seviyeli istatistiksel öznitelikler, enerji olmak üzere 12 adet öznitelik hesaplanmıştır. Oluşturulan öz nitelikleri birleştirilerek veri seti oluşturulmuştur. Bu sayede her gönüllü son İMF'leri artık olarak sayılan İMF'ler çıkarıldığında 48 (4x12) ile 72 (6x12) arasında değişen sayıda yeni formda sinyalle veri setinde temsil edilmiştir.

Oluşturulan veri kümesinin k-parçalı çapraz doğrulama metoduyla 5 parçaya bölünmüştür. Her parçanın eşit sayıda astımlı ve sağlıklı gönüllünün oskültasyon sesi alınmasına dikkat edilmiştir. Eğitim kümesi için kullanılan özniteliklerin gönüllüler için rastgele seçilmiş ve gönüllülerin geri kalanları test kümesi olarak ele alınmıştır. Bu sayede tüm oskültasyon sesleri performansın hesaplanması sırasında hem test hem de eğitim aşamasında kullanılarak popülasyona dâhil edilmiştir. Çalışmada Matlab paket programında oluşturulmuş DİA kullanılmıştır. Modellenen DİA 2 gizli katmandan meydana gelmektedir. Gizli katmalarda kullanılan nöron sayıları 60 ile 300 nöron arasında deneysel olarak seçilmiş önerilen DİA sınıflandırıcı da çapraz doğrulama ile elde edilen en yüksek başarım paylaşılmıştır. Önerilen DİA ilk gizli katmanında 130 adet nöron, ikinci katmanında ise 190 adet nöron kullanılmıştır. Sistem sağlıklı ve astım olarak iki çıkış vermektedir. DİA'nın ilk kısmında kısıtlanmış Boltzmann makineleri kullanılarak denetimsiz öğrenmeye tabi tutularak nöronların ağırlıkları belirlenmiş ve belirlenen ağırlıkların denetimli öğrenme kısmında sigmoid fonksiyon bazlı geriye yönelim ağı kullanılarak optimizasyonu sağlanmıştır. DİA'nın her denetimli öğrenme sırasında global değere erişmek için 100 iterasyon uygulanmış ve öğrenme katsayısı 0.2 olarak belirlenmiştir. Rastgele seçilen eğitim kümelerine ait hem 12 adet öznitelik 12 nöronlu giriş katmanına sahip DİA yapısı kullanılarak eğitilmiştir. Eğitim sonrası test edilen farklı nöron sayısına sahip modeller ile farklı başarım performansları göstermiştir (Tablo I). En yüksek performansa sahip derin öğrenme modelinin yapısına ilk gizli katmanında 160 nöron, ikinci gizli katmanında 130 nöron bulunan derin modelin performansını belirlemek için çapraz doğrulama tablosundan (Tablo II) hassasiyet, belirlilik ve genel başarım değerleri hesaplanmıştır.

| DBN Katman ve Nöron Yapısı | Hassasiyet | Belirlilik | Genel Başarım |
|---|---|---|---|
| 200-310 | **%91.06** | %81.91 | %83.33 |
| 150-300 | %87.85 | %77.43 | %82.06 |
| 160-130 | %85.83 | %77.11 | **%84.61** |

**Tablo I**. Farklı DBN yapılarıyla elde edilen sınıflandırıcı performansları

|  | Astım | Sağlıklı |
|---|---|---|
| **Astım** | 509 | 60 |
| **Sağlıklı** | 84 | 283 |

**Tablo II**. Astım Sınıflandırma modelinin çapraz tablosu

DİA modelinin çapraz sınıflandırma tablosuna göre elde edilen sonuçlar kendi içerisinde değerlendirildiğinde modele tüm öznitelikler dâhil edildiğinde astımlı bireylerden kaydedilen akciğer oskültasyon sesleri ile sağlıklı bireylerden kaydedilen akciğer oskültasyon sesleri %84.61 genel başarım, %85.83 hassasiyet ve %77.11 belirlilik ile oldukça yüksek performansla ayırt edilebilmektedir.

## IV. TARTIŞMA

Literatürde astımlı bireyden elde edilen akciğer oskültasyon seslerinin sağlıklı akciğer seslerinden ayrıştırılmasını sağlama üzerine çalışan ve hırıltılı seslerin teşhisi üzerine yapan çalışmalar mevcuttur.

Bunlar içerisinde farklı frekans analiz yöntemleri, farklı sensör ve medikal cihazlarla elde edilen verileri kullanan, farklı oskültasyon merkezlerinden alınan seslerin analizini yapan çalışmaların karşılaştırılması oldukça güçtür. Bu çalışmalar içerisinden göğüs duvarına yerleştirilmiş 14 mikrofon ve pnömotakograf kullanarak teager enerji operatörü, renyi entropi değerleri ve kısa-zamanlı Fourier dönüşümünden elde edilen yeni formdaki sinyallerin basıklık değerleri ile f50/f90 klinik ölçüm değerini kullanarak çevrimiçi sağlık hizmetlerinde hırıltının ölçülebildiği ispatlanmış [12], oskültasyon seslerine uygulanan Q faktörü ayarlanabilme özelliğine sahip rasyonel yayılımlı dalgacık dönüşümünden elde edilen yeni formdaki sinyallerin güç dağılımı ve istatistiksel öznitelikleri destek vektör makineleri kullanarak ortalama %95.17 genel başarımla hırıltıları ve normal sesleri ayırt edebildiği ispatlanmıştır [14]. Sağ-üst akciğer bölgesinden alınan oskültasyon sesine uygulanan Hilbert dönüşümü, hızlı Fourier dönüşümü sonrası elde edilen ortalama genlik, ortalama akım, ortalama güç değerlerinin istatistiksel regresyon analizleriyle astımlı bireyler ve normal bireylerden elde edilen oskültasyon sesler arasındaki bağlantı değerlendirilmiştir [11].

| Çalışma | Kullanılan Yöntem | Hassasiyet | Belirlilik | Genel Başarım |
|---|---|---|---|---|
| **Guntupalli vd.** [23] | Titreşim Tepki Görüntüleme | %91 | %82 | %85 |
| **Waitmann vd.** [24] | Ortalama Güç spektrumu | %62 | %85 | %73 |
| **Bahoura** [25] | Mel frekans filtreleri + Dalgacık Dönüşümü | %68.8 - %94.6 | %80.3 - %91.9 | %76.27 - %92.2 |
| **Bu çalışma** | Hilbert-Huang Dönüşümü | %85.83 | %77.11 | %84.61 |

**Tablo III**. Astım sınıflandırması yapan benzer çalışmalar ve sınıflandırıcı performansları

Titreşim tepki görüntüleme sistemiyle akciğerin hırıltılı bölgelerinin tespitini yapan [23], akciğer seslerinden elde edilen ortalama güç spektrum değerlerini kullanarak astım sınıflandırma yapan [24], akciğer seslerine uygulanan mel frekans filtreleri ve dalgacık dönüşümlerine Gaussian karışım modelini uygulayarak farklı değerlendirmelerle astımlı hastalardan kaydedilmiş akciğer seslerinin sağlıklı sınıfından ayrıştırılmasını sağlayan çalışmalar [25] yüksek sınıflandırıcı başarımlarıyla literatürde yer almaktadır (Tablo III). Bu çalışmalara alternatif ve ek öznitelik olarak kullanılabilecek HHD ile elde edilen İMF özniteliklerden elde edilen istatistiksel ve yüksek seviyeli istatistik öznitelikleri derin öğrenme algoritması olarak sıklıkla tercih edilen iki katmanlı DİA sınıflandırıcısıyla %84.61 genel başarımla yüksek bir ayrıştırma kabiliyetine sahip olduğu deneyimlenmiştir. Bu çalışmayla frekans bazında öznitelik çıkarma metotlarına ek frekans-zaman bazında elde edilecek özniteliklerin DİA ile bütünleşik bir sistemde astımlı ve sağlıklı bireylerden alınan göğüs ve sırttan alınan akciğer oskültasyon seslerinin analizleri sonucunda yüksek performansla ayrıştırılabildiği, frekans domeninde olduğu kadar zaman domeninde de anlamlı öznitelikler çıkarılabileceği gösterilmiştir.

## TEŞEKKÜR

## KAYNAKÇA